\documentclass[prc,aps,floatfix,nofootinbib,twocolumn,superscriptaddress,preprintnumbers]{revtex4-2} 

\usepackage{graphicx}
\usepackage{color}
\usepackage{enumerate}
\usepackage{amssymb}
\usepackage{amsmath}
\usepackage{amsfonts}
\usepackage{bbm}
\usepackage{dsfont}
\usepackage{url}
\usepackage{ulem}
\usepackage[colorlinks=true,allcolors=blue,breaklinks]{hyperref}

\renewcommand{\vec}[1]{\mbox{\boldmath $#1$}}

\begin{document}

\preprint{KUNS-3068, YITP-25-120, J-PARC-TH-0320}

\title{
Probing surface vibration of spherical nuclei in relativistic heavy-ion collisions}

\author{Kouichi Hagino}
\email{hagino.kouichi.5m@kyoto-u.ac.jp}
\affiliation{ 
Department of Physics, Kyoto University, Kyoto 606-8502,  Japan} 
\affiliation{Institute for Liberal Arts and Sciences, Kyoto University, Kyoto 606-8501, Japan}
\affiliation{ 
RIKEN Nishina Center for Accelerator-based Science, RIKEN, Wako 351-0198, Japan
}

\author{Masakiyo Kitazawa}
\email{kitazawa@yukawa.kyoto-u.ac.jp}
\affiliation{ 
Yukawa Institute for Theoretical Physics, Kyoto University, Kyoto 606-8502, Japan}
\affiliation{ 
J-PARC Branch, KEK Theory Center, Institute of Particle and Nuclear Studies, KEK, Tokai, Ibaraki 319-1106, Japan}

\begin{abstract}
There has been increasing interest in recent years in 
using relativistic heavy-ion collisions to 
probe nuclear structure, such as static nuclear deformation. 
Here we discuss the role of quantum zero-point fluctuations of the surface vibration of spherical nuclei in relativistic heavy-ion collisions. 
To this end, we employ an approach to describe the vibration in the space-fixed frame, which has been well established in the field of low-energy heavy-ion fusion reactions. We particularly consider the quadrupole vibration of $^{58}$Ni 
in $^{58}$Ni+$^{58}$Ni reaction  
and the octupole vibration of $^{208}$Pb in $^{208}$Pb+$^{208}$Pb reaction.
We show that the surface vibration leads to comparable eccentricity parameters to those for static deformation, while they give significantly different distributions of the initial states,
suggesting the importance of the proper treatment of the surface vibration in heavy-ion collisions. 
We perform similar analysis also for triaxial deformation and gamma-soft vibration.  
\end{abstract}

\maketitle

{\it Introduction.}
Nuclear deformation is one of the most important concepts in nuclear structure physics 
\cite{RingSchuck}. 
When a nucleus is permanently deformed, it exhibits a characteristic rotational band, in which 
the excitation energies of a state with spin $I$ is proportional to $I(I+1)$. 
It also shows enhanced electromagnetic transition strengths as well as large quadrupole 
moments. Moreover, it has also been well known that nuclear deformation significantly 
affects low-energy nuclear reactions \cite{Hagino2012,Hagino2022}. In particular, heavy-ion fusion 
reactions at energies around the Coulomb barrier are sensitive to nuclear 
deformation \cite{Hagino2012,Hagino2022,Dasgupta1998,Leigh1995,Leigh1993,Wei1991}, and there have been 
many attempts to determine deformation parameters of a nucleus, especially the sign of hexadecapole deformation parameter $\beta_4$ \cite{Leigh1995,Leigh1993,Gupta2020,Gupta2023}, 
using the fusion barrier distribution \cite{Dasgupta1998,Timmers1995}. 

In recent years, there has been increasing interest in probing nuclear deformation in relativistic heavy-ion 
collisions (RelHIC)~\cite{giacalone2020,giacalone2021,bally2022,jia2022,jia2022-2,zhang2022,ryssens2023,jia2023,xu2024,jia2024,star2024,star2025,zhao2024,wang2024,zhang2025,fortier2025,li2025}. 
The key idea of this approach is that different deformations of colliding nuclei lead to different distributions of the initial states on the transverse plane. Through the hydrodynamic response of the quark-gluon plasma, they turn into the event-by-event distribution of the collective flow, which is experimentally observable and carries information about the nuclear deformation.
An important point here is that the energy (time) scale of heavy-ion reactions in RelHIC is much larger (shorter) than that of nuclear motions. One can thus adopt a simple picture of adiabatic approximation, i.e., the initial nuclear configuration is regarded to be frozen during a collision.
This approach has been applied not only to the quadrupole deformation $\beta_2$ but also the hexadecapole deformation $\beta_4$ \cite{xu2024,ryssens2023}, the triaxial quadrupole deformation $\gamma$ \cite{zhao2024,bally2022,jia2022-2}, 
the octupole deformation $\beta_3$ \cite{zhang2025}, and the alpha cluster structure \cite{cao2023,giacalone2025,liu2025}. 
For example, the STAR Collaboration has  reported in Ref.~\cite{star2024,star2025} the deformation parameters $\beta_2=-0.286\pm0.025$ and $\gamma=8.7\pm4.8^\circ$ for $^{238}$U.
See Ref. \cite{jia2024} for a recent review (see, however, also  Ref.~\cite{Dobaczewski:2025rdi} for a recent comment on these 
activities). 

Whereas these studies have mainly focused on the \textit{static} deformation of deformed nuclei so far, the initial states of RelHIC can also be affected by the quantum zero-point fluctuations of the deformation, whose importance is well recognized in the low-energy sub-barrier fusion reactions~\cite{Hagino2012,Hagino2022,Dasgupta1998,balantekin1998,stefanini1995,morton1999,esbensen1987}. 
In RelHIC, a recent paper by Xu, {\it et al}.\ has shown that a puzzle related to 
the anisotropic flow 
observed in ultra-central $^{208}$Pb+$^{208}$Pb collisions~\cite{Allice1,Allice2,Allice3,Allice4,Allice5} can be resolved 
to a large extent by the octupole surface vibration of the $^{208}$Pb nucleus~\cite{xu2025}. 
The fluctuation of the triaxial deformation parameter $\gamma$ has also been discussed in Ref.~\cite{zhao2024}. 

In the present Letter, we discuss the effect of the quantum fluctuations of deformation parameters on RelHIC focusing on the surface vibrations of \textit{spherical} nuclei.
To describe the vibrations, we employ the harmonic-oscillator model in the \textit{space-fixed} frame~\cite{Hagino2012,Hagino2022}. 
This treatment can describe 
harmonic vibrations in all the directions on equal footing,
and connect their magnitudes to the experimental data on the electronic transition probability.
Using this method, we explore the eccentricity parameters and mean-square area of the initial states of RelHIC based on the liquid-drop model with a complete overlap~\cite{jia2022} for ${}^{58}$Ni and ${}^{208}$Pb.
Comparing these results with the case of the static deformation, we demonstrate that these treatments lead to qualitatively different distributions from each other. The difference is especially highlighted in the skewness and kurtosis of the eccentricity parameters, suggesting their discrimination through the measurement of higher-order correlations of flow observables in the ultra-central collisions. Similar analysis is also performed for the triaxial deformation parameter $\gamma$.

{\it Surface vibration.}
To describe the deformation of a nucleus, we use a deformed Woods-Saxon density 
\begin{equation}
    \rho(\vec{r})=\frac{\rho_0}{1+e^{(r-R(\theta,\phi))/a}}, 
    \label{eq:Woods-Saxon}
\end{equation}
for the nucleon density in the {\it space-fixed} frame with the polar coordinates $(r,\theta,\phi)$, where $\rho_0$ and $a$ are the central density and the diffuseness parameter, respectively.  
Here,  $R(\theta,\phi)$ is the angle dependent radius given by 
\begin{equation}
     R(\theta,\phi)=R_0\left(1-\frac{1}{4\pi}\sum_{\lambda,\mu}|\alpha_{\lambda\mu}|^2
     +\sum_{\lambda,\mu}\alpha_{\lambda\mu}Y^*_{\lambda\mu}(\hat{\vec{r}})\right),
     \label{eq:radius}
\end{equation}
with the deformation parameters in the space-fixed frame  $\alpha_{\lambda\mu}$ ($-\lambda\le\mu\le\lambda$) satisfying $\alpha^*_{\lambda\mu}=(-1)^\mu\alpha_{\lambda,-\mu}$, the radius parameter $R_0$, and $\hat{\vec r}=\vec r/r$. The second term in the parenthesis in Eq.~\eqref{eq:radius} is due to the 
volume conservation~\cite{RingSchuck}.

In the harmonic oscillator model for spherical nuclei, the classical Hamiltonian for the deformation parameters $\alpha_{\lambda\mu}$ reads~\cite{RingSchuck}
\begin{align}
    H &=\frac{1}{2}\sum_{\lambda,\mu}\left(B_\lambda|\dot{\alpha}_{\lambda\mu}|^2+C_\lambda|\alpha_{\lambda\mu}|^2\right)
    \notag \\
    &=\frac{1}{2}\sum_{\lambda,\mu}\left(B_\lambda\,\dot{\tilde \alpha}_{\lambda\mu}^2+C_\lambda\,\tilde{\alpha}_{\lambda\mu}^2\right) ,
    \label{eq:Hcl}
\end{align}
where the dot denotes the time derivative, and 
$B_\lambda$ and $C_\lambda$ are the inertia and the stiffness parameters of the oscillator, respectively. 
On the second line, we have introduced the real coordinates 
\begin{equation}
    \tilde{\alpha}_{\lambda\mu}\equiv \frac{1}{\sqrt{2}}(\alpha_{\lambda\mu}+\alpha^*_{\lambda\mu}),~~~
    \tilde{\alpha}_{\lambda,-\mu}\equiv \frac{1}{\sqrt{2}i}(\alpha_{\lambda\mu}-\alpha^*_{\lambda\mu}),  
\end{equation}
for $\mu>0$ and $\tilde{\alpha}_{\lambda0}=\alpha_{\lambda0}$ 
for $\mu=0$. 
Equation~\eqref{eq:Hcl} shows that there exist $2\lambda+1$ independent harmonic oscillators for a given $\lambda$ with 
$\{\tilde{\alpha}_{\lambda\mu}\}$ being the coordinates. 

In quantum mechanics, the ground state of these harmonic oscillators has the zero-point fluctuation with $\langle \tilde{\alpha}_{\lambda\mu}\rangle=0$ and $\langle \tilde{\alpha}_{\lambda\mu}^2\rangle=\sigma_\lambda^2$, 
where the amplitude of the zero-point motion is $\sigma_\lambda=\sqrt{\hbar/(2B_\lambda\omega_\lambda)}$ with 
$\omega_\lambda=\sqrt{C_\lambda/B_\lambda}$. 
We thus have
\begin{equation}
    \left\langle\sum_\mu|\alpha_{\lambda\mu}|^2\right\rangle
    =\left\langle\sum_\mu \tilde{\alpha}_{\lambda\mu}^2\right\rangle
    =(2\lambda+1)\sigma_\lambda^2\equiv(\beta_\lambda)^2. 
    \label{eq:beta}
\end{equation}
In low-energy heavy-ion fusion reactions, $\beta_\lambda$ is often referred to as the (dynamical) deformation parameter~\cite{Hagino2012,Hagino2022}. 
Its value can be estimated from a measured electronic transition 
probability $B(E\lambda)\uparrow$ from the ground state of a spherical nucleus to an excited state as~\cite{Hagino2012,Hagino2022},
\begin{equation}
    \beta_\lambda=\frac{4\pi}{3ZR_0^\lambda}\,\sqrt{\frac{B(E\lambda)\uparrow}{e^2}}, 
\end{equation}
where $Z$ is the atomic number of the nucleus. 
From the measured $B(E2)\uparrow=0.0695~e^2\,{\rm b}^2$ for the transition from 
the ground state to the first 2$^+$ state at 1.45\,MeV in 
$^{58}$Ni \cite{raman2001}, $\beta_2$ is estimated to be 
$\beta_2(^{58}{\rm Ni})=0.218$ with $R_0=1.1\times 58^{1/3}$\,fm. 
In the case of $^{208}$Pb, the measured $B(E3)\uparrow=0.608~e^2\,{\rm b}^3$ \cite{NNDC}
for the transition from the ground state to the first 3$^-$ state at 2.61\,MeV 
yields $\beta_3(^{208}{\rm Pb})=0.144$ with $R_0=1.1\times 208^{1/3}$\,fm. 
It has been well recognized that excitations of those vibrational states considerably 
affect heavy-ion fusion reactions of e.g., $^{58}$Ni+$^{58}$Ni and 
$^{16}$O+$^{208}$Pb, at energies around the Coulomb barrier \cite{Hagino2012,Hagino2022,esbensen1987,stefanini1995,morton1999,yao2016}. 

It is instructive to compare the deformation parameters in Eq.~\eqref{eq:radius} to those for static deformation. 
The static deformation is usually expressed in the \textit{body-fixed} frame with an appropriate choice of the principal axes~\cite{Greiner}. 
The deformation parameters $a_{\lambda\mu}$ in this frame are related to those in the space-fixed one as~\cite{RingSchuck}, 
\begin{equation}
    \alpha_{\lambda\mu}=\sum_{\mu'}D^\lambda_{\mu'\mu}(\Omega)a_{\lambda\mu'}, 
    \label{eq:defpara}
\end{equation}
where $D^\lambda_{\mu'\mu}$ is the Wigner $D$ function and $\Omega$ is the Euler angle between the two frames. 
For $\lambda=2$, the quadrupole deformation parameters in the body-fixed frame 
are usually parametrized as
\begin{equation}
a_{20}=\beta_2\cos\gamma, 
\quad
a_{2,\pm1}=0,
\quad 
a_{2,\pm2}=\frac{\beta_2}{\sqrt2}\sin\gamma.
\label{eq:gamma}
\end{equation}
Substituting Eq.~\eqref{eq:gamma} into Eq.~\eqref{eq:defpara}, the deformation parameters in the space-fixed system read 
\begin{equation}
    \alpha_{2\mu}=D^2_{0\mu}(\Omega)\beta_2\cos\gamma +\frac{1}{\sqrt{2}}\left(D^2_{2\mu}(\Omega)+D^2_{-2\mu}(\Omega)\right)
    \beta_2\sin\gamma. 
\end{equation}
For an axially-symmetric shape, only $a_{\lambda0}=\beta_\lambda$ has a nonzero value in $a_{\lambda\mu}$, and 
Eq. (\ref{eq:defpara}) is transformed to 
\begin{equation}
    \alpha_{\lambda\mu}=\beta_\lambda D^\lambda_{0\mu}(\Omega).  
     \label{eq:defpara-axial}
\end{equation}

We mention that the surface vibration can be described in the body-fixed frame using $a_{\lambda\mu}$ by transforming the Hamiltonian~\eqref{eq:Hcl} via Eq.~\eqref{eq:defpara}. However, the Hamiltonian is no longer separable in this case.  
One the other hand, the rotational symmetry ensures the separable 
structure of Eq.~\eqref{eq:Hcl} in the space-fixed frame, that is an advantage to use the latter for spherical nuclei.

{\it Initial geometry of RelHIC.}
Next, we investigate the effect of the surface vibration in RelHIC.
For high collision energies, the time scale of heavy-ion reactions is significantly shorter than that of the vibrational degrees of freedom. This justifies the use of the adiabatic approximation, where 
$\{\tilde{\alpha}_{\lambda\mu}\}$ are regarded as constants during the reaction. The transverse structure of the initial states of RelHIC is determined simply by the overlap of colliding nuclei, whose shapes are specified by 
$\{\tilde{\alpha}_{\lambda\mu}\}$ distributed according to Eq.~\eqref{eq:beta}. This property is contrasted to the low-energy heavy-ion reactions, where vibrational degrees of freedom are not slow for most nuclei and the adiabatic approximation~\cite{esbensen1981} is not adequate. In this case, one needs to solve the coupled-channels equations as they are~\cite{Hagino2012,Hagino2022}.

In Refs.~\cite{giacalone2020,jia2022-2}, it has been pointed out that the initial conditions for the ultra-central collisions are approximately given by the completely overlapped collisions. 
In the present study, for a simple illustration of the importance of the surface vibration in RelHIC, we follow Ref.~\cite{jia2022} and assume that the transverse distribution of the energy density in the initial state is proportional to 
\begin{equation}
    \rho^{(z)}(\vec{r}_\perp)=\int^\infty_{-\infty}dz\,\rho(\vec{r}) ,
\end{equation}
where $\vec{r}_\perp=(x,y)$ is the transverse vector with the $z$-axis chosen along the beam direction.

In the following, we focus on the eccentricity parameters and the inverse mean-square area
\begin{equation}
\epsilon_n
=
-\frac{\langle\!\langle (x-iy)^n\rangle\!\rangle}{\langle\!\langle (x^2+y^2)^{n/2} \rangle\!\rangle}, 
\qquad
d = \frac1{\sqrt{\langle\!\langle x^2 \rangle\!\rangle \langle\!\langle y^2 \rangle\!\rangle}} ,
\label{eq:ed}
\end{equation}
respectively, of the initial conditions, in which 
$\langle\!\langle\cdots\rangle\!\rangle$ means the average with respect 
to $\rho^{(z)}(\vec{r}_\perp)$. 
It is well established that $|\epsilon_n|$ and $d$ are strongly correlated with the anisotropic flows $v_n$ and the mean transverse momentum $\overline{p_T}$, respectively, in the final state~\cite{Niemi:2012aj}. Using the correlations, the distribution of $\epsilon_n$ and $d$ in the initial states is encoded in the event-by-event distribution of $v_n$ and $\overline{p_T}$ that are experimentally observable~\cite{jia2022-2}.

In the harmonic-oscillator model~\eqref{eq:Hcl}, the distributions of $\tilde{\alpha}_{\lambda\mu}$ obey the Gaussian function $P_{\rm G}(x)\propto e^{-x^2/(2\sigma_\lambda^2)}$, which are independent with one another. The expectation value of 
$|\epsilon_n|^2$ over collision events,
for example, is thus evaluated as
\begin{equation}
    \langle |\epsilon_n|^2\rangle 
    =\int \left(\prod_{\lambda,\mu} d\tilde{\alpha}_{\lambda\mu}\,P_{\rm G}(\tilde{\alpha}_{\lambda\mu})\right)|\epsilon_n(\{\tilde{\alpha}_{\lambda\mu}\})|^2,
    \label{eq:ecc_vib}
\end{equation}
where $\epsilon_n(\{\tilde{\alpha}_{\lambda\mu}\})$ is the eccentricity parameter in Eq.~(\ref{eq:ed}) for a given set of $\{\tilde{\alpha}_{\lambda\mu}\}$.
On the other hand, for the static deformation with axial symmetry, the expectation value is given by the integral over the Euler angle $\Omega$ as~\cite{jia2022}
\begin{equation}
    \langle |\epsilon_n|^2\rangle 
    =\int \frac{d\Omega}{8\pi^2}\, |\epsilon_n(\Omega)|^2 ,
        \label{eq:ecc_rot}
\end{equation}
where $\epsilon_n(\Omega)$ for a given $\Omega$ is evaluated by substituting Eq.~\eqref{eq:defpara-axial} into Eq.~\eqref{eq:radius} in the space-fixed frame. 

\begin{figure}[tb] 
\begin{center} 
\includegraphics[width=0.8\columnwidth]{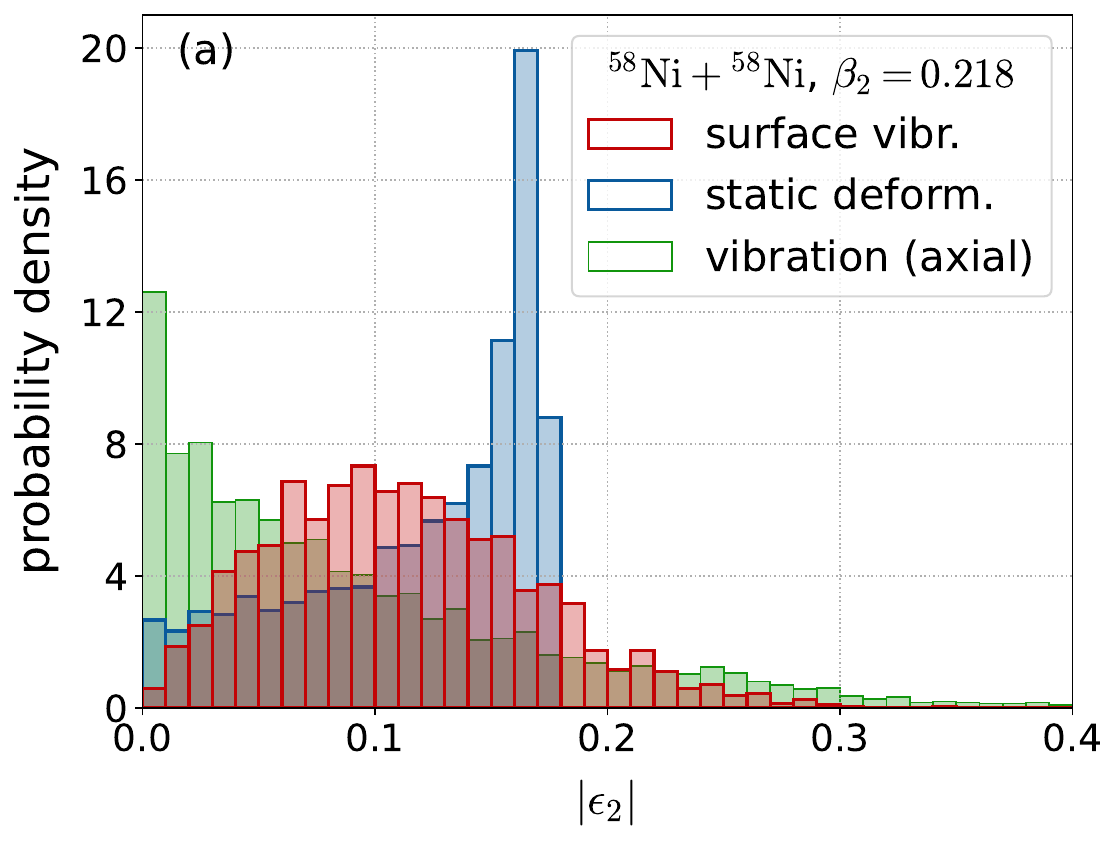} 
\includegraphics[width=0.8\columnwidth]{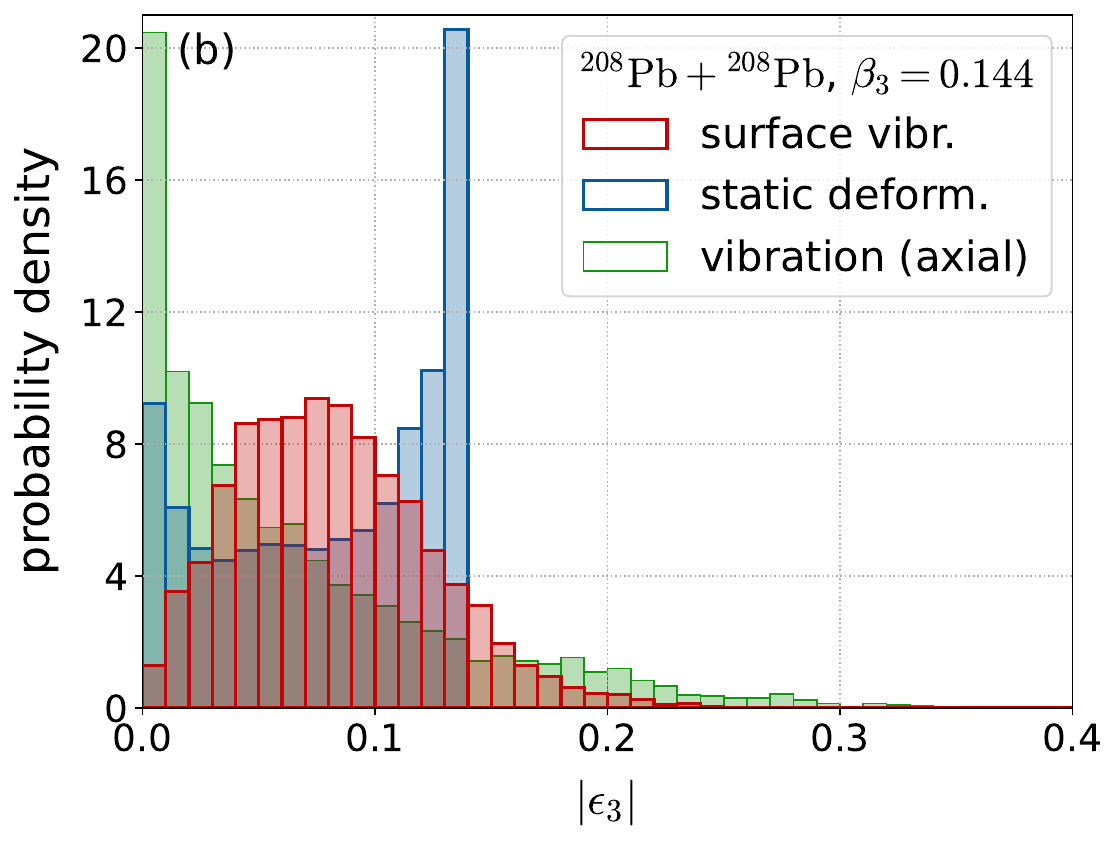} 
\caption{(a) Probability densities of the 
quadrupole eccentricity parameter $|\epsilon_2|$ for the $^{58}$Ni+$^{58}$Ni collision. 
The red, blue, and green colors denote the results for the surface vibration (SV), static deformation (SD), and the surface vibration with axial (SV-A). 
(b) The same as (a), but for the octupole eccentricity $|\epsilon_3|$ for 
the $^{208}$Pb+$^{208}$Pb collision. 
}
\label{fig:eccentricity}
\end{center} 
\end{figure}

{\it Distribution of $\epsilon_n$.}
Let us now numerically evaluate the distribution of the eccentricity parameters for spherical nuclei with the surface vibration (SV) for $^{58}$Ni+$^{58}$Ni and $^{208}$Pb+$^{208}$Pb collisions, and compare them with those for a static deformation (SD). 
For this purpose, we generate $\{\tilde{\alpha}_{\lambda\mu}\}$ randomly according to the probability distribution in Eq.~\eqref{eq:ecc_vib} for SV, while for SD only $\Omega$ is randomly sampled with fixed $\beta_\lambda$ for the axial shape.
Figure~\ref{fig:eccentricity} shows the probability densities of the quadrupole eccentricity parameters $|\epsilon_2|$ for the $^{58}$Ni+$^{58}$Ni collision (upper) and the octupole eccentricity parameters 
$|\epsilon_3|$ for the $^{208}$Pb+$^{208}$Pb collision (lower) with 3000 random samples.
To calculate the eccentricities, we use the deformed Woods-Saxon density (\ref{eq:Woods-Saxon}) with $a=0.55\,$fm and $R_0=1.1A^{1/3}\,$fm, where $A$ is the mass number of a nucleus.
For the $^{58}$Ni+$^{58}$Ni ($^{208}$Pb+$^{208}$Pb) collision, 
we take only the $\lambda=2$ ($\lambda=3$) components in $\alpha_{\lambda\mu}$ with $\beta_2=0.218$ ($\beta_3=0.144$). 
The results for SV are shown by the red color, while the SD results with the same static deformation parameters $\beta_\lambda$ are shown by the blue color. 
From the figure one finds that the surface vibration leads to as large $|\epsilon_n|$ as the static deformation~\cite{xu2025}. 
Moreover, SV and SD lead to completely different distributions of the eccentricity parameters for each collision system.
The difference can be clarified in the mean, standard deviation, skewness, and kurtosis of these distributions~\cite{Asakawa:2015ybt} as shown in Table~\ref{tab:val}. In particular, the skewness has opposite signs for SV and SD.

\begin{table}[]
    \centering
    \begin{tabular}{l|l|cccc}
         \hline
         \hline
         &&  mean & std.\ dev. & skewness & kurtosis \\
         \hline
         ${}^{58}$Ni, $|\epsilon_2|$ 
        & SV & 0.112(1) & 0.0554(7) & 0.49(3) & -0.02(11) \\
        & SD & 0.119(1) & 0.0500(5) & -0.79(3) & -0.62(6) \\
        & SV-A & 0.090(1) & 0.0816(13)& 1.22(4) & 1.12(20) \\
         \hline
         ${}^{208}$Pb, $|\epsilon_3|$ 
        & SV & 0.0822(8) & 0.0416(5) & 0.55(4) & 0.15(11) \\
        & SD & 0.0821(8) & 0.0461(4) & -0.38(3) & -1.29(3) \\
        & SV-A & 0.0650(12) & 0.0649(11) & 1.35(5) &  1.49(22) \\
          \hline
         \hline
    \end{tabular}
    \caption{The mean, the standard deviation, the skewness, and the kurtosis of the eccentricity parameter $|\epsilon_2|^2$ ($|\epsilon_3|^2$) of ${}^{58}$Ni$+{}^{58}$Ni (${}^{208}$Pb$+{}^{208}$Pb) collision. 
    See Ref.~\cite{Asakawa:2015ybt} for the definition of these quantities.
    The results for surface vibration (SV), static deformation (SD), and surface vibration with axial only (SV-A) in Eq.~\eqref{eq:ecc_vib2} are} shown with the statistical errors.
    \label{tab:val}
\end{table}

In treating the surface vibration, one sometimes uses a similar formula to Eq.~(\ref{eq:ecc_rot}) 
but takes account of the fluctuation of the axial deformation parameter, $\beta_\lambda$, only~\cite{xu2025,denisov2006}, so that the expectation value of $|\epsilon_n|^2$ reads
\begin{equation}
    \langle |\epsilon_n|^2\rangle 
    \approx\int P_{\rm G}(\beta_\lambda)d\beta_\lambda\int \frac{d\Omega}{8\pi^2}\, |\epsilon_n(\beta_\lambda,\Omega)|^2 .
    \label{eq:ecc_vib2}
\end{equation}
This treatment would be equivalent to Eq.~\eqref{eq:ecc_vib} if the fluctuations of the non-axial deformation 
parameters were also taken into account in an appropriate manner, even though the non-axial 
degrees of freedom are not usually taken into account. 
The probability densities in this treatment (SV-A) are shown by the green color in Fig.~\ref{fig:eccentricity}.
One can see that the results are qualitatively different from both SV and SD. The statistical quantities in Table~\ref{tab:val} also highlight the difference.

The large values of $|\epsilon_n|$ found in the above results for SV imply that the surface vibration substantially affects the initial states of RelHIC, and its appropriate incorporation into the dynamical modeling of RelHIC is crucial~\cite{xu2025}. The significant difference between SV, SD, and SV-A also suggests that they are distinguishable experimentally. In particular, the behaviors of the skewness and kurtosis indicate that the higher-order cumulants of the anisotropic flow $v_n$ are sensitive to it.
In this connection, it would be interesting to investigate not only the mean value but also the the distributions of the elliptic flow 
shown e.g. in Fig.~2 in Ref.~\cite{star2024}. 

\begin{figure}[tb] 
\begin{center} 
\includegraphics[width=0.9\columnwidth]{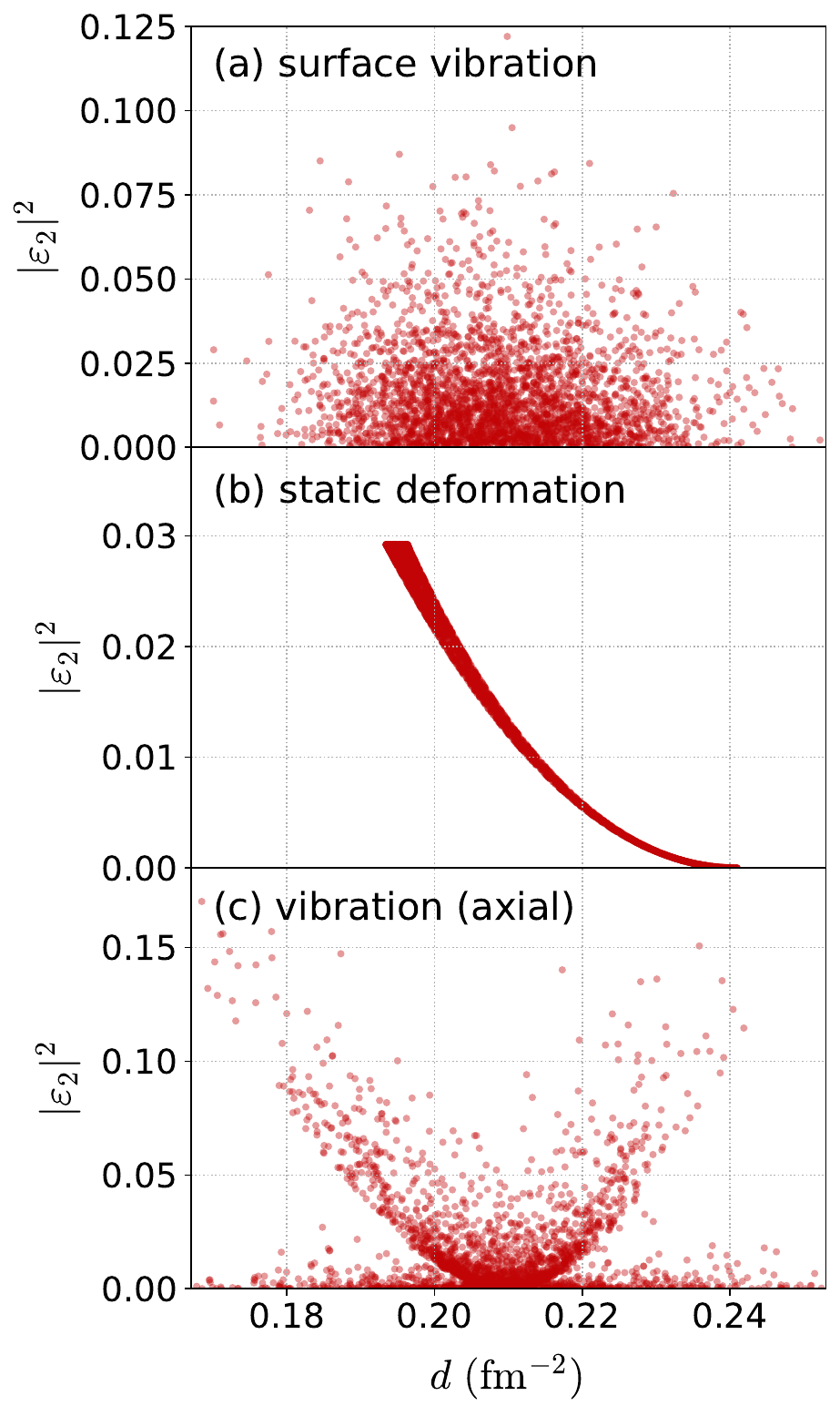} 
\caption{The distributions of the quadrupole eccentricity parameters $\epsilon_2$ for the 
the $^{58}$Ni+$^{58}$Ni collision with the deformation parameter of $\beta_2=0.218$. 
These are plotted as a function of the 
inverse of the mean square area, 
$d\equiv 1/\sqrt{\langle\!\langle x^2\rangle\!\rangle
\langle\!\langle y^2\rangle\!\rangle}$. 
The top, the middle, and the bottom panels show the distributions of $|\epsilon_2|$ for the dynamical deformation, 
the static deformation, and the fluctuation of the axial deformation parameter, respectively, which correspond 
to the three lines in the upper panel of Fig. \ref{fig:eccentricity}. 
}
\label{fig:distribution}
\end{center} 
\end{figure} 

{\it $\epsilon_2$--$d$ correlation.}
To gain further insights into the difference of the three cases, in Fig.~\ref{fig:distribution} we plot the distribution of the initial states for each case for ${}^{58}$Ni$+{}^{58}$Ni on the $|\epsilon_2|^2$--$d$ plane~\footnote{We find that the distribution remains qualitatively the same except for the direction of the slope, even if the horizontal axis is replaced with the root mean square radius 
$r_\perp\equiv\sqrt{\langle\!\langle x^2+y^2\rangle\!\rangle}$ instead of $d$.}. 
The figure shows that the distributions are significantly different for the three cases. For SV (top panel), $|\epsilon_2|$ and $d$ have almost no correlation. In contrast, for the case of SD (middle panel),
they are strongly correlated with each other. 
The treatment with SV-A (bottom panel) has a characteristic behavior different both from SV and SD.

The behaviors for SV and SD can be understood easily in the space-fixed frame as follows. Using the sharp-cut density $a=0$ in Eq.~\eqref{eq:Woods-Saxon} and to the first order of $\alpha_{\lambda\mu}$, $|\epsilon_2|$ and $d$ are analytically calculated as~\cite{jia2022}
\begin{equation}
    |\epsilon_2|=\sqrt{\frac{15}{2\pi}}\,|\alpha_{22}|,  
    \qquad
    d=
    \frac{5}{R_0^2}\left(1+\sqrt{\frac{5}{4\pi}}\alpha_{20}\right). 
    \label{eq:e2-d}
\end{equation}
Because $\alpha_{20}$ and $\alpha_{22}$ are independent in Eq.~\eqref{eq:Hcl}, one immediately finds from Eq.~\eqref{eq:e2-d} that $|\epsilon_2|$ and $d$ are uncorrelated in SV within this approximation. On the other hand, for SD $\alpha_{20}$ and $\alpha_{22}$ have one-to-one correspondence through Eq.~\eqref{eq:defpara-axial}. 
These simple arguments are another advantage of the space-fixed frame. 

From the strong correlation between $\epsilon_2$ and $v_2$, as well as $d$ and $\overline{p_T}$, it is suggested from Fig. \ref{fig:distribution} that SV, SD, and SV-A give significantly different consequences of the experimental results, especially in the correlations of $v_2$ and $\overline{p_T}$~\cite{jia2022-2,bozek2012,zhang2025}. It is also interesting to observe the event-by-event distribution on the $v_2$--$\overline{p_T}$ plane experimentally besides the correlation functions, as it is clearly different among SV, SD, and SV-A.

{\it Triaxiality.} 
Finally, in connection with these results let us briefly discuss the fluctuation in the triaxial deformation~\cite{zhao2024}. 
The triaxial quadrupole deformation of a nucleus is usually parametrized by $\beta_2$ and $\gamma$ in Eq.~\eqref{eq:gamma}~\footnote{A parametrization of 
non-axial deformation may not be trivial for higher order deformation with $\lambda > 2$ 
\cite{hamamoto1991,nonaxial,wexler1999,yoshida2025}.}.
In Ref.~\cite{zhao2024}, the difference between the static and dynamical triaxial deformations has been investigated for ${}^{129}$Xe$+{}^{129}$Xe collision, assuming the $\gamma=\pi/6$ for the former case, while $\gamma$ is randomly distributed in the latter.

\begin{figure}[tb] 
\begin{center} 
\includegraphics[width=0.9\columnwidth]{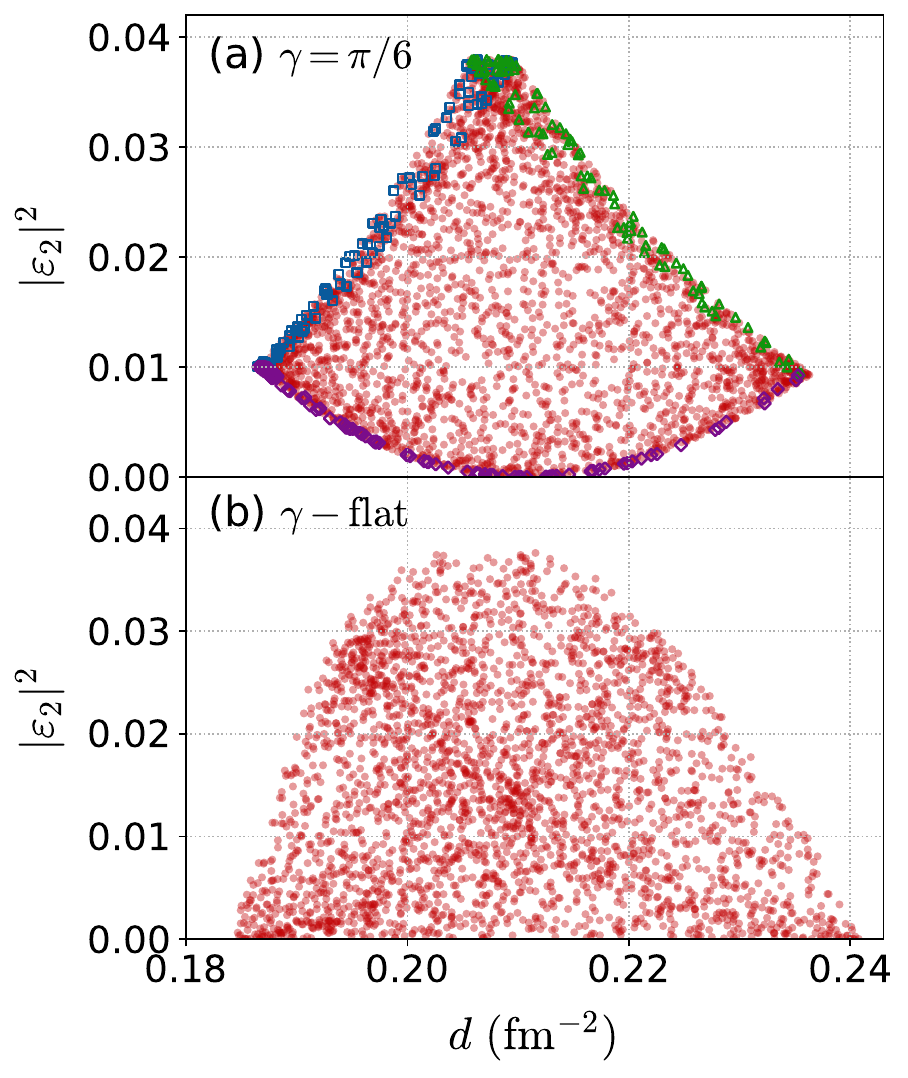} 
\caption{Same as Fig. \ref{fig:distribution}, but for the triaxial deformation. 
The upper panel corresponds to a static triaxial deformation with $\gamma=\pi/6$, 
while the lower panel is obtained by assuming a flat distribution of $\gamma$ in the range from 0 to 
$\pi/3$. The value of $\beta_2$ is set to be $\beta_2=0.218$. In the upper panel, the eccentricity parameters 
for specific Euler angles, $\theta=\pi/2,\phi=0$, and $\phi=\pi/2$, are denoted by the blue squares, the green 
triangles, and the purple diamonds, respectively. 
}
\label{fig:triaxial}
\end{center} 
\end{figure} 

Figure~\ref{fig:triaxial} shows the distributions of the initial states on the $|\epsilon_2|^2$--$d$ plane obtained for ${}^{58}$Ni with $\beta_2=0.218$ and the same 
parameter set for the density distribution
as in Fig.~\ref{fig:distribution}. The upper panel corresponds to the case of a rigid triaxial deformation with $\gamma=\pi/6$, 
while the lower panel shows the results with a flat distribution of $\gamma$ between 0 and $\pi/3$~\cite{zhao2024}. 
The mean values of $|\epsilon_2|^2$ are 0.0157 and 0.0151 
for the upper and the lower panels, respectively. 
The upper panel also shows the eccentricity parameters for specific values of the Euler angles, 
that is, $\theta=\pi/2$ (the blue squares), $\phi=0$ (the green triangles), and $\phi=\pi/2$ (the purple diamonds), 
which provide the edges of the distribution. 

One can see that the distributions are considerably different between these two cases, suggesting that they can be distinguished experimentally, even though the mean value of the eccentricity parameter is similar to each other. 
This is in a similar situation as Fig.~\ref{fig:distribution}. Whereas the higher-order correlations of $v_2$ and $\overline{p_T}$ have been investigated in Ref.~\cite{zhao2024}, 
we advocate the event-by-event distribution on the $v_2$--$\overline{p_T}$ plane
for a clearer distinction between a static and a dynamical 
triaxial deformation. 

{\it Summary.}
We have discussed the effects of the quantum zero-point fluctuations of colliding nuclei on the initial states of relativistic heavy-ion collisions in the ultra-central 
region. In particular, we have investigated the role of surface vibration of spherical nuclei. To deal with the vibrational degrees of freedom, we have used a formalism based on the space-fixed frame, which is widely used in low-energy heavy-ion fusion reactions. 
This formalism treats all the deformation parameters 
$\alpha_{\lambda\mu}$ equally 
for the same multipolarity, $\lambda$, in contrast to the formalism in the body-fixed frame where the axial and the non-axial deformation parameters are separately treated. 
Applying the formalism to the $^{58}$Ni+$^{58}$Ni and $^{208}$Pb+$^{208}$Pb collisions, we have 
shown that the surface vibrations lead to comparable eccentricities to those for static deformations. 
Moreover, we have shown that the dynamical and the static deformations lead to considerably different 
correlation plots in the two-dimensional plane of the eccentricity and the inverse of the 
mean square area, indicating that 
those two can be experimentally distinguished with such plots. 
We have also shown that this is the case for triaxial deformation, that is, the difference 
between a static triaxial deformation and a flat distribution in the $\gamma$-direction. 

In this Letter, we have used a simple model based on the complete overlap of colliding nuclei~\cite{jia2022-2} to calculate the eccentricity parameters of the initial states in RelHIC. Although we believe that this treatment is sufficient to reveal the qualitative importance of the surface vibration, more sophisticated analyses based on dynamical models~\cite{zhao2024,xu2024,zhang2025,xu2025,fortier2025} are necessary for a better reproduction of the real experiments such as imperfect correlations between the initial eccentricities and the anisotropic flow in the final state, and effects of the off-central collisions. These investigations are left for future study.

{\it Acknowledgments.}
We thank Jiangyong Jia and Chunjian Zhang 
for useful discussions during the workshop ``Intersection of nuclear structure and high-energy nuclear collisions: 2025'' held at Fudan University, Shanghai, China, May 12--22, 2025. 
This work was supported in part by
JSPS KAKENHI Grant Numbers JP23K03414 and JP24K07049.

\bibliography{ref}
\end{document}